\documentclass[conference]{IEEEtran}

\ifdefined\labelindent
\else
\newlength{\labelindent}
\fi

\usepackage[utf8]{inputenc}
\usepackage[T1]{fontenc}
\usepackage[pdftex]{graphicx} 
\graphicspath{{../pdf/}{../jpeg/}} 
\DeclareGraphicsExtensions{.pdf,.jpeg,.png} 

\usepackage[cmex10]{amsmath}
\usepackage{amssymb} 
\usepackage{array}
\usepackage{url} 
\usepackage{url} 

\usepackage{booktabs}
\usepackage{multirow}
\usepackage[nocompress]{cite} 
\usepackage{xcolor} 
\usepackage{tikz} 

\usepackage{tabularx}

\usepackage{algorithm}
\usepackage{algpseudocode}

\usepackage{geometry}
\definecolor{lime}{HTML}{A6CE39} 
\DeclareRobustCommand{\orcidicon}{%
    \begin{tikzpicture}
        \draw[lime, fill=lime] (0,0)
            circle [radius=0.16]
            node[white] {\tiny ID};
        \draw[white, fill=white] (-0.0625,0.095)
            circle [radius=0.007];
    \end{tikzpicture}%
    \hspace{0mm}
}
\newcommand{\orcidauthor}[2]{%
    #1\,\orcidicon
}

\begin{document}
\title{
A MUSIC-Based Adaptive Beamforming Approach for Mobile Jamming Mitigation in 5G Networks
} 
\author{ 
    \IEEEauthorblockN{
        \orcidauthor{Olivia Holguin}{0009-0001-5858-3814},
        \orcidauthor{Rachel Donati}{0009-0000-9790-5030},
        \orcidauthor{Seyed bagher Hashemi Natanzi}{0000-0003-1524-8669-0000}\IEEEauthorrefmark{1}, 
        \orcidauthor{Bo Tang}{0000-0001-5708-766X}
    }
    \IEEEauthorblockA{
        \IEEEauthorrefmark{1}Department of Electrical and Computer Engineering, Worcester Polytechnic Institute, Worcester, MA, USA\\ 
        Email: \{oholguin, rddonati, snatanzi, btang1\}@wpi.edu
    }
}
\maketitle

\begin{abstract}
Mobile jammers pose a critical threat to 5G networks, particularly in military communications. This paper investigates an anti-jamming framework that enhances a strong adaptive beamforming baseline comprising Multiple Signal Classification (MUSIC) for Direction-of-Arrival (DoA) estimation and Minimum Variance Distortionless Response (MVDR) for interference suppression with a lightweight machine learning (ML) model for predictive error correction. Extensive simulations in a realistic highway scenario demonstrate that the integrated system achieves a high DoA estimation accuracy of up to 99.8\% and an average Signal-to-Noise Ratio (SNR) improvement of 9.58 dB. Analysis reveals that the MUSIC-MVDR baseline alone accounts for the vast majority of this performance gain (9.46 dB), indicating that the primary benefit of the simple ML model lies in correcting outlier estimates rather than providing a substantial systemic SNR increase. The framework's computational efficiency validates the effectiveness of the core beamforming approach and highlights the critical trade-off between ML model complexity and practical performance gains for securing 5G communications in contested environments.
\end{abstract}

\IEEEoverridecommandlockouts
\vspace{1mm}

\IEEEoverridecommandlockouts
\vspace{1mm}

\IEEEpeerreviewmaketitle

\section{Introduction}
\label{sec:Introduction}

The widespread adoption of 5G networks has transformed wireless communications, enabling high-speed and low-latency connectivity for a range of critical applications. However, these networks are increasingly vulnerable to sophisticated jamming attacks, particularly from mobile jammers that dynamically change their position to disrupt reliable communication \cite{9733393}. Beamforming, a fundamental technology in 5G, enhances signal quality and mitigates interference but faces significant challenges against mobile jammers, as traditional anti-jamming methods rely on static assumptions unsuitable for dynamic interference sources.

Conventional interference mitigation strategies, such as fixed beamforming patterns, are effective against stationary jammers but fail to adapt to the mobility of jammers, which degrade network performance by introducing intentional interference across the spectrum \cite{9733393}. The combination of Multiple Signal Classification (MUSIC) for Direction-of-Arrival (DoA) estimation and Minimum Variance Distortionless Response (MVDR) for interference suppression is a well-established and powerful approach for adaptive beamforming. However, the performance of this combination can still be challenged by highly dynamic scenarios where DoA estimation errors may occur. This presents an opportunity to investigate whether lightweight computational methods can enhance the robustness of this framework without incurring significant processing overhead \cite{10149863}.

\begin{figure}[htbp]
    \centering
    \includegraphics[width=0.9\columnwidth]{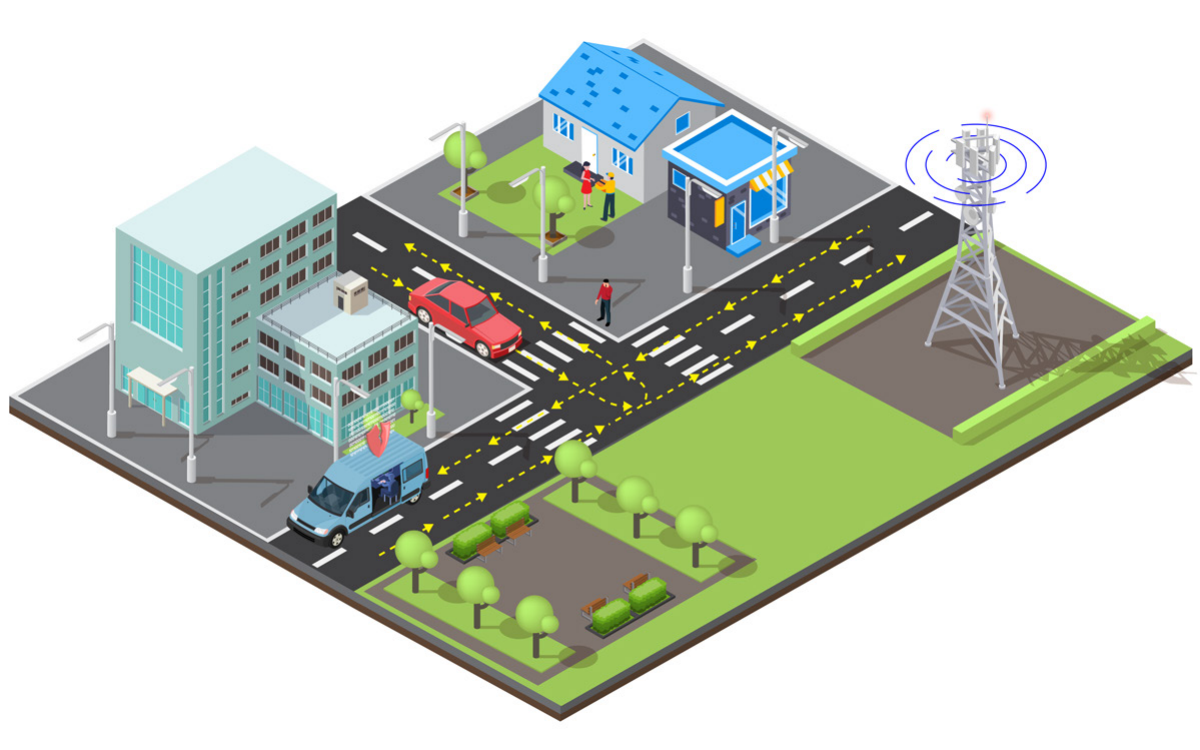}
    \caption{Simulated highway scenario illustrating the interaction between a stationary transmitter (Tx), a mobile receiver (Rx), and a mobile jammer (Jammer), designed to evaluate dynamic anti-jamming performance.}
    \label{fig:scenario_setup}
\end{figure}

This paper explores an enhanced anti-jamming framework designed for mobile jammers in 5G networks, as depicted in Fig.~\ref{fig:scenario_setup}. The framework builds upon the MUSIC-MVDR baseline and integrates a lightweight machine learning (ML) model to improve DoA estimation accuracy, particularly in cases where the primary MUSIC algorithm yields high error. The key objectives are: (1) to implement a robust MUSIC-MVDR baseline for mobile jammer mitigation, (2) to assess the potential of a simple linear regression model to correct DoA estimation errors with minimal computational cost, and (3) to provide a comprehensive performance analysis quantifying the trade-offs between performance gain and complexity. The results demonstrate robust performance in mobile jamming scenarios while maintaining computational efficiency suitable for real-time applications.

The remainder of the paper is structured as follows: Section~\ref{Background} reviews background and related work. Section~\ref{Proposed_Framework} describes the proposed MUSIC-MVDR-ML framework. Section~\ref{Experiment} presents experimental results, discussion, and limitations. Finally, Section~\ref{conclusion} concludes the paper and outlines future research directions.

\section{Background and Related Work}
\label{Background}

Beamforming, essential for enhancing signal quality and mitigating interference in modern wireless communications like 5G, has evolved from conventional fixed-weight methods to adaptive solutions \cite{benesty_brief_2021}. While conventional beamforming offers simplicity, its fixed patterns struggle against dynamic interference, particularly mobile jammers. Adaptive techniques like MVDR and Linear-Constrained Minimum Variance (LCMV) adjust parameters based on the signal environment.

MVDR beamforming optimizes array weights to maximize sensitivity towards a desired DoA while suppressing other sources \cite{darzi_null_2014}. However, its performance degrades if the desired DoA is inaccurate. To address this, MVDR is commonly coupled with high-resolution DoA estimation algorithms like MUSIC, which uses eigenspace decomposition to accurately identify signal directions even in multi-source scenarios \cite{1143830, 10366174}. LCMV also minimizes interference but imposes linear constraints to preserve the target signal \cite{7471643}. A practical limitation is that LCMV is often restricted to a uniform linear array (ULA), whereas MVDR offers more flexibility, adapting well to the uniform rectangular array (URA) used in this study, making it a suitable choice for investigation \cite{10366174}.

Traditional anti-jamming strategies focusing on spatial filtering are effective against stationary jammers but fail when confronting mobile ones that dynamically change positions \cite{10811542}. Recent research has explored integrating machine learning to address this limitation. While complex models like deep neural networks are being investigated, this study emphasizes the trade-off between performance and efficiency by employing a simple, interpretable ML model. At the network layer, multipath communications and secret sharing schemes also enhance resilience by distributing data across multiple paths to counter jamming attacks \cite{10773801, 10773704}. The proposed approach complements these strategies by focusing on the physical layer, where the combination of MUSIC's accurate DoA estimation, MVDR's adaptive nulling, and supplementary ML support presents a practical method for mobile interference mitigation.

Finally, accurately evaluating performance through signal-to-noise ratio (SNR) in multi-antenna systems facing mobile jammers presents challenges. Standard SNR calculations may not adequately handle dynamically changing interference or the specifics of signal combination after beamforming, necessitating careful consideration of the calculation methodology, as addressed in the proposed framework. Limitations of current DoA estimation with mobile sources, real-time adaptation challenges, and the need for better ML integration highlight the necessity for the comprehensive approach developed in this study.

\section{Proposed Framework}
\label{Proposed_Framework}

The proposed framework integrates MUSIC-based DoA estimation, MVDR beamforming, and ML enhancement for DoA estimation support. A $2 \times 2$ URA receives signals from a desired transmitter and a mobile jammer. The received signal $x(t)$ at the array is modeled as:
\begin{equation}
    x(t) = s(t)a(\theta_s,\phi_s) + j(t)a(\theta_j,\phi_j) + n(t)
\end{equation}
where $s(t)$ is the desired signal, $j(t)$ is the jamming signal, $n(t)$ is additive white Gaussian noise, and $a(\theta,\phi)$ is the array steering vector corresponding to azimuth angle $\theta$ and elevation angle $\phi$. Subscripts $s$ and $j$ denote the desired signal and jammer, respectively.

\subsection{Anti-Jamming System}

\vspace{0.5em}
\noindent\textbf{MUSIC-Enhanced DoA Estimation:}
The MUSIC algorithm estimates angles of arrival by exploiting the eigenstructure of the received signal's covariance matrix $R$ \cite{1143830}:
\begin{equation}
   R = E\{x(t)x^H(t)\}
\end{equation}
Eigendecomposition of $R$ separates the signal subspace ($U_s$) from the noise subspace ($U_n$) \cite{matlab_music}:
\begin{equation}
    R = U_s\Lambda_sU_s^H + U_n\Lambda_nU_n^H
\end{equation}
High-resolution DoA estimation is achieved by scanning over a grid of possible angles, in this study from -90$^{\circ}$ to 90$^{\circ}$, and evaluating the MUSIC pseudospectrum. Peaks in the pseudospectrum indicate the estimated DoA of incoming signals.

\vspace{0.5em}
\noindent\textbf{Adaptive MVDR Beamforming:}
Beamforming is a spatial filtering strategy where weight geometries and gain tapering methods manipulate transmitter and receiver antennas. A conventional beamformer remains static, while an adaptive beamformer responds to environmental changes, allowing the antenna array to adjust dynamically. The MVDR beamformer computes optimal weights $w_{\text{opt}}$ to minimize output power while preserving the desired signal gain in the direction $\theta_s$ \cite{matlab_mvdr}:
\begin{equation}
    w_{\text{opt}} = \frac{R^{-1}a(\theta_s,\phi_s)}{a^H(\theta_s,\phi_s)R^{-1}a(\theta_s,\phi_s)}
\end{equation}
This adaptively creates nulls in the direction of interference sources, such as the mobile jammer. The weights $w_{\text{opt}}$ are then applied to the antenna array for beam shaping.

\begin{figure}[htbp]
    \centering
    \includegraphics[width=0.85\columnwidth]{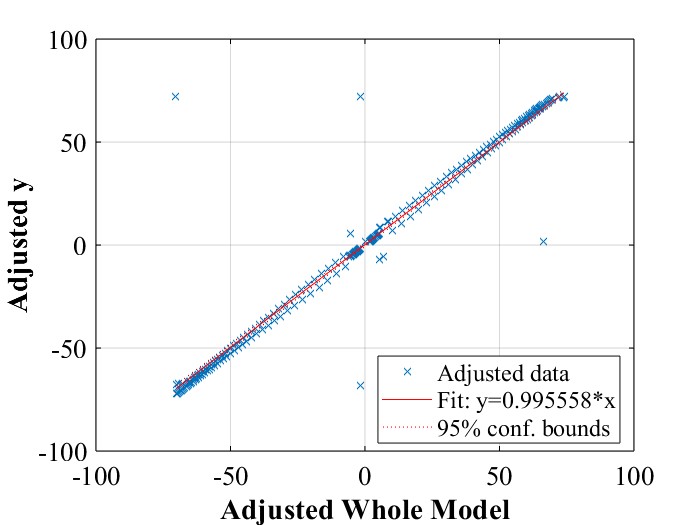} 
    \caption{Optimized linear regression model showing the relationship between predicted and actual azimuth angles with 95\% confidence bounds.}
    \label{fig:ml_plot}
\end{figure}

\vspace{0.5em}
\noindent\textbf{Machine Learning Integration:}
To enhance DoA estimation accuracy when initial MUSIC estimates exceed a predefined error threshold, a linear regression ML model is employed. Linear regression is chosen for its simplicity and interpretability. The model is trained using MATLAB's fitlm \cite{matlab_ml}, with input features as sequences of three consecutive ground truth azimuth angles and the output as the subsequent ground truth azimuth angle. Over 500 DoA data points from simulated mobility patterns are used as training data (Fig.~\ref{fig:ml_plot}). The model predicts the next azimuth angle $\hat{\theta}_{t+1}$ based on the past $k$ measurements $(\theta_t, \theta_{t-1}, ..., \theta_{t-k})$:
\begin{equation}
\begin{aligned}
\hat{\theta}_{t+1} &= \beta_{0} + \sum_{i=1}^k \beta_i\cdot\theta_{t-i}
\end{aligned}
\label{eq:ml_prediction}
\end{equation}
While $k=3$ is used in these experiments, the model can be generalized to include more historical measurements. The statistical significance of the $\theta_{t-2}$ coefficient (p-value $\approx 3.4\times10^{-76}$) confirms strong temporal dependency, supporting the predictive capability of the model.

\subsection{Performance Quantification}
\vspace{0.5em}
\noindent\textbf{SNR Calculation:}
Performance evaluation uses an SNR calculation method that averages antenna signals across the $M \times N$ array elements to prevent artificial gain inflation and ensure consistency between simulation and potential implementation. The combined signal is $x_2(t) = \frac{1}{N+M} \sum_{i=1}^{N+M} |x(t)_{i}|$, and signal and noise powers are calculated over signal-present ($a_1$ to $b_1$) and signal-absent ($a_0$ to $b_0$) regions:
\begin{align}
    \label{eq:sig_power} \text{Signal Power} &= \frac{1}{b_1 - a_1 + 1} \sum_{i=a_1}^{b_1} |x_2(i)| \\
    \label{eq:noise_power} \text{Noise Power} &= \frac{1}{b_0 - a_0 + 1} \sum_{i=a_0}^{b_0} |x_2(i)| \\
    \label{eq:snr_db} \text{SNR (dB)} &= 20 \log_{10} \left( \frac{\text{Signal Power}}{\text{Noise Power}} \right)
\end{align}
SNR improvement is computed as:
\begin{equation}
    \label{eq:snr_improvement} \text{SNR}_{\text{improvement}} = \text{SNR}_{\text{after}} - \text{SNR}_{\text{before}}
\end{equation}

\begin{figure}[htbp]
    \centering
    \includegraphics[width=0.9\columnwidth]{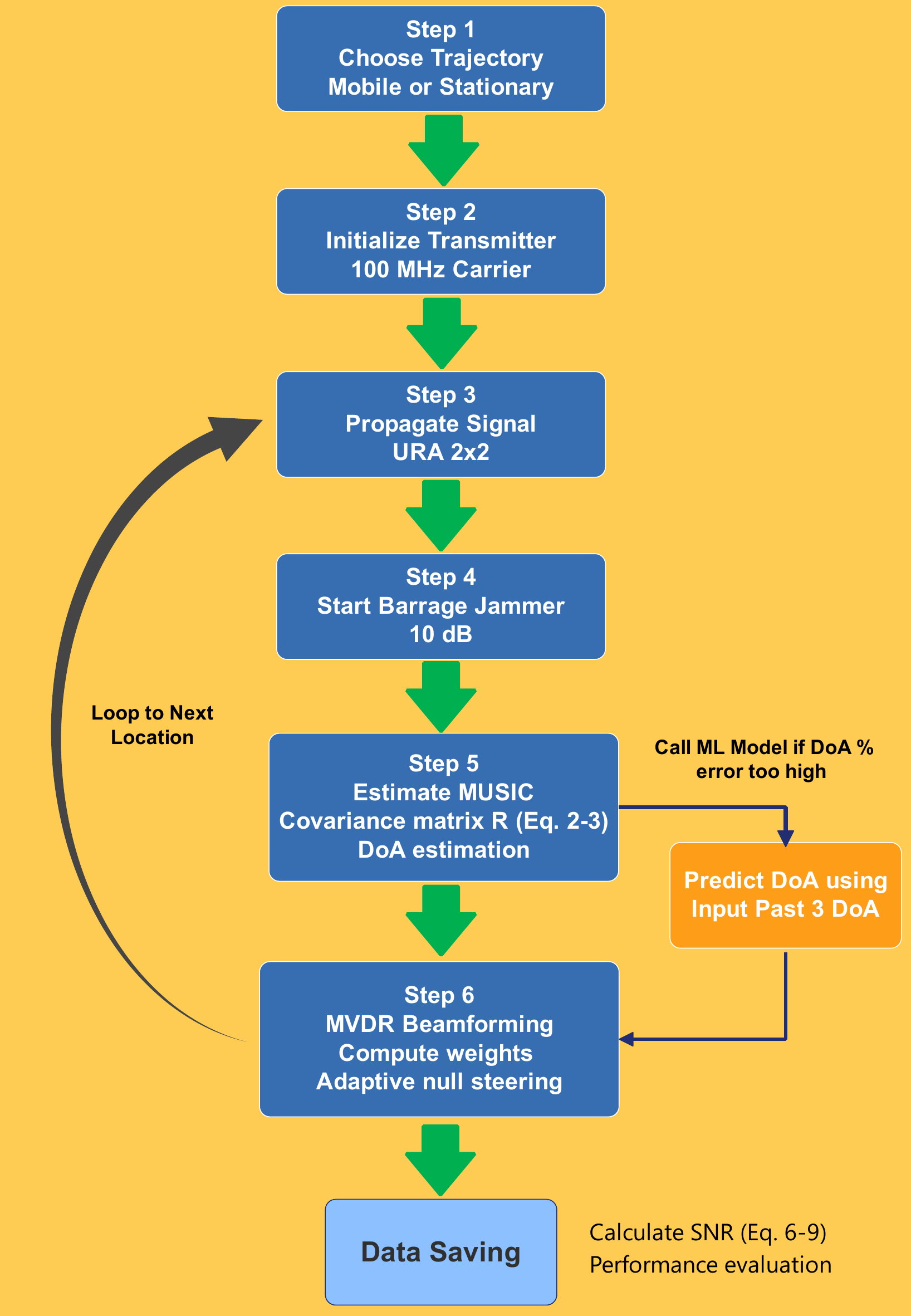} 
    \caption{Block diagram of the proposed MUSIC-MVDR-ML framework showing the step-by-step mitigation process.}
    \label{fig:flowchart}
\end{figure}

\vspace{0.5em}
\noindent\textbf{Mitigation Algorithm:}
Fig.~\ref{fig:flowchart} illustrates Algorithm~\ref{alg:jammer}, which outlines the simulation framework for mitigating mobile jamming by integrating MUSIC-based DoA estimation, MVDR beamforming, and the ML prediction model. The algorithm iteratively collects signals, estimates DoA using MUSIC (optionally refined by the ML model when errors exceed the threshold), computes MVDR weights, applies beamforming, and evaluates performance based on SNR improvement. The system can adjust parameters or retrain the ML model if necessary to improve performance. This integrated approach provides robust protection against mobile jammers while maintaining computational efficiency.

\begin{algorithm}[htbp]
\caption{Mobile Jammer Mitigation}
\label{alg:jammer}
\begin{algorithmic}[1]
\Require Received signal stream $x(t)$, array parameters, ML model
\Ensure Optimized MVDR weights, enhanced SNR performance
\vspace{0.5em}
\State \textbf{Initialize:} Array parameters, ML model, and DoA threshold $\epsilon$
\While{System is operational}
    \State Collect current signals $x(t)$
    \State Compute covariance matrix $R \gets E[x(t)x^H(t)]$
    \State Estimate DoA angles $\{\theta_s,\, \phi_s\} \gets \mathrm{MUSIC}(R)$
    \If{$\text{DoA Error} > \epsilon$}
        \State Predict $\theta_s$ as $\hat{\theta}_{t+1}$ using ML model
    \EndIf
    \State Update ML model with latest measurements for next prediction
    
    \State Compute MVDR weights $w_{\mathrm{opt}} \gets \frac{R^{-1} a(\theta_s,\phi_s)}{a^H(\theta_s,\phi_s)\, R^{-1}\, a(\theta_s,\phi_s)}$
    \State Apply beamforming with $w_{\mathrm{opt}}$ and measure $\text{SNR}_{\text{after}}$
    \State Calculate $\text{SNR}_{\text{improvement}}$

\EndWhile
\end{algorithmic}
\end{algorithm}

\section{Experiment}
\label{Experiment}

\subsection{Setup}
The proposed framework was evaluated through extensive MATLAB simulations emulating a realistic highway scenario (Fig.~\ref{fig:scenario_setup}). The scenario involved a stationary transmitter (cell tower), a mobile receiver (red car), and a mobile jammer (blue van) moving along predefined paths. The receiver and jammer could begin from any of the three cardinal directions (in the figure, the receiver began on the west side). Trajectory options included North-to-South, North-to-West, South-to-North, South-to-West, West-to-North, and West-to-South. Key simulation parameters are as follows:

The carrier frequency of 100 MHz was selected to simplify the simulation and reduce computational complexity. The MUSIC-MVDR-ML framework is frequency-agnostic and can be adapted to standard 5G bands, such as 3.5 GHz, in future implementations. A signal sampling rate of 1 kHz was used for baseband processing after downconversion, sufficient for the simulated scenario.

\begin{itemize}
    \item Jammer: Mobile or Stationary
    \item Carrier frequency: 100 MHz
    \item Signal sampling rate: 1 kHz
    \item Antenna array: $2 \times 2$ URA
    \item Jammer power: 10 dB
\end{itemize}

\subsection{Results}
\vspace{0.5em}
\noindent\textbf{DoA Estimation and ML Performance:}
The integrated MUSIC algorithm demonstrated high DoA estimation accuracy (up to \textbf{99.8\%}) in the presence of a dynamic jammer across the tested scenarios. This accuracy was achieved using a hierarchical error threshold system with tolerance levels of 15\%, 10\%, and 7.5\% for combined azimuth/elevation estimation errors, calculated as the percentage deviation from the total angle span. The ML model activates when MUSIC fails three consecutive estimation attempts, using linear regression based on the previous three DoA measurements.

\begin{figure}[htbp]
    \centering
    \includegraphics[width=0.75\columnwidth]{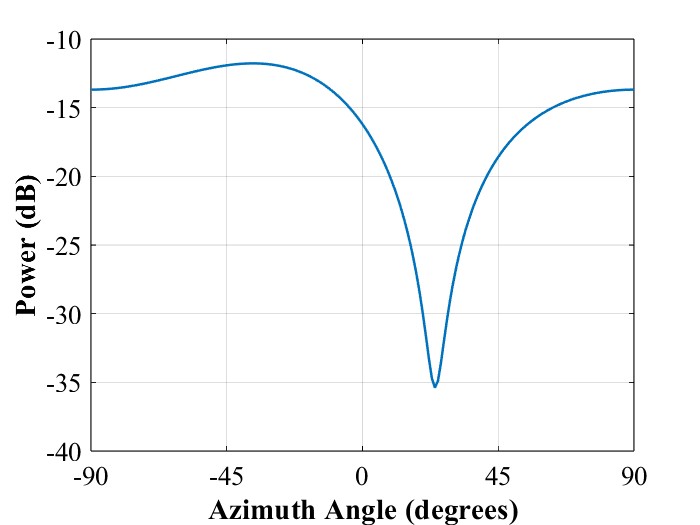} 
    \caption{Example MVDR beam pattern demonstrating effective null steering towards the jammer direction (e.g., at 30$^{\circ}$) while maintaining gain towards the desired signal (e.g., at -40$^{\circ}$).}
    \label{fig:beam_pattern}
\end{figure}

Fig.~\ref{fig:beam_pattern} shows an example beam pattern with the estimated DoA indicated by its peak at -40$^{\circ}$. Machine learning support aids the estimation when error exceeds a threshold.

\vspace{0.5em}
\noindent\textbf{Adaptive Beamforming:}
The adaptive MVDR beamformer effectively steered nulls towards the jammer direction while preserving the desired signal. Analysis of beam patterns (e.g., Fig.~\ref{fig:beam_pattern}) confirmed improved directionality and interference suppression compared to initial states. The system mitigated jamming interference along both linear and diagonal highway trajectories.

\begin{figure}[htbp]
    \centering
    \includegraphics[width=0.75\columnwidth]{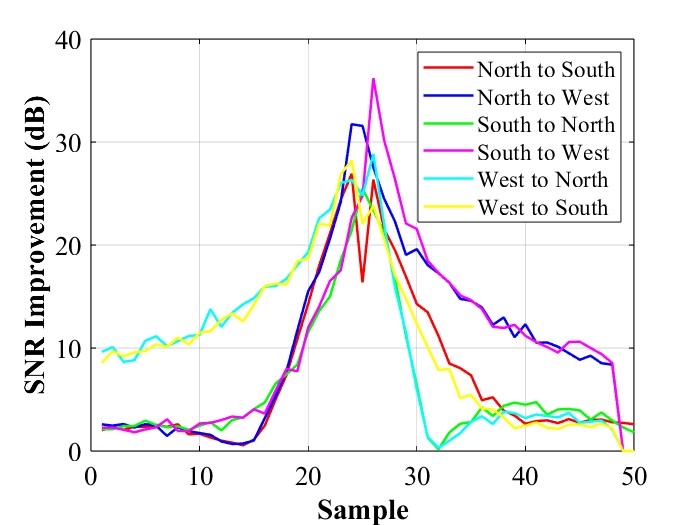} 
    \caption{SNR improvement achieved by the proposed MUSIC-MVDR-ML framework along a representative highway trajectory, showcasing consistent enhancement despite jammer mobility.}
    \label{fig:snr_improvement_trends} 
\end{figure}

\vspace{0.5em}
\noindent\textbf{SNR Enhancement:}
Using the defined SNR calculation method, which averages combined antenna signals, the framework achieved significant improvement. Across six diverse highway trajectories, the system yielded an average SNR improvement of \textbf{9.58 dB} post-MVDR beamforming. While the improvement varied depending on relative positions and paths (Fig.~\ref{fig:snr_improvement_trends}), all trajectories exhibited a spike in SNR around sample 25, corresponding to the highway intersection. This enhancement demonstrates the framework's anti-jamming capability.

\vspace{0.5em}
\noindent\textbf{Computational Efficiency:}
The framework is suitable for real-time applications, with average processing times per collection as follows: DoA estimation (MUSIC) $\approx$ \textbf{41.9 ms}, MVDR weight computation $\approx$ \textbf{5.2 ms}, and ML prediction $\approx$ \textbf{16.9 ms}\footnote{\texttt{github.com/CLIS-WPI/mobile-jamming-mitigation}.}.

\vspace{0.5em}
\noindent\textbf{Baseline Comparison:}
Adaptive beamforming and machine learning were compared across MATLAB simulations. Performance metrics for SNR improvement, DoA accuracy, and processing time for three methods are summarized in Table~\ref{tab:algorithm_comparison}.

\begin{table}[htbp]
\renewcommand{\arraystretch}{1.2}
\centering
\caption{Detailed Algorithm Comparison}
\label{tab:algorithm_comparison}
\begin{tabular}{lccc}
\toprule
\textbf{Algorithm} & \textbf{SNR (dB)} & \textbf{Accuracy (\%)} & \textbf{Type} \\
\midrule
Fixed Beamforming & 1.90 & 88.0 & Static \\
MUSIC-MVDR & 9.46 & 99.3 & Adaptive \\
MUSIC-MVDR-ML & 9.58 & 99.8 & ML-Enhanced \\
\bottomrule
\end{tabular}
\end{table}

\subsection{Discussion}
The experimental results demonstrate the effectiveness of the adaptive beamforming framework. A critical analysis of components and limitations provides further insights.

\vspace{0.5em}
\noindent\textbf{Analysis of ML Component Impact:}
The baseline comparison shows that the MUSIC-MVDR framework alone achieves an average SNR improvement of \textbf{9.46 dB}, while adding the ML model increases this to \textbf{9.58 dB}. This marginal gain of 0.12 dB indicates that, in the simulated highway trajectories, the MUSIC algorithm is highly robust, leaving limited room for improvement by a simple predictive model.

The superiority of MUSIC-MVDR-ML stems from three key factors: (1) MUSIC provides sub-degree DoA resolution through eigenspace decomposition, (2) MVDR creates precise adaptive nulls using optimal weight computation $w_{\text{opt}} = R^{-1}a(\theta_s,\phi_s) / [a^H(\theta_s,\phi_s)R^{-1}a(\theta_s,\phi_s)]$, and (3) the ML model serves as an outlier correction mechanism when MUSIC estimation quality degrades. The modest ML improvement validates that the MUSIC-MVDR baseline already achieves near-optimal performance in single-jammer scenarios.

The linear regression model primarily acts as a corrective mechanism for outlier DoA estimates, improving accuracy from 99.3\% to 99.8\% while exerting minimal impact on SNR. This highlights a crucial trade-off between pursuing higher performance with complex ML models and maintaining a computationally efficient framework that ensures baseline effectiveness and reliability.

\vspace{0.5em}
\noindent\textbf{Limitations:}
Despite robust performance, several limitations remain. The relatively modest impact of the current ML model suggests that future work should explore more sophisticated architectures, such as Recurrent Neural Networks (RNNs) or Long Short-Term Memory (LSTM) networks, which can better capture temporal dependencies in jammer mobility patterns and potentially provide greater gains. Such models, however, require larger datasets and rigorous validation techniques (e.g., cross-validation) to prevent overfitting.

Additionally, performance against multiple simultaneous jammers has not yet been investigated, as only a single jammer was simulated. Multi-jammer scenarios would require modeling multiple jammer objects and channels in each collection. Finally, the serial simulation loop, in which signal propagation occurs at every collection before processing, introduces latency. Future implementations should consider multi-threading or parallel processing to achieve true real-time operation.

The controlled simulation environment with predictable highway mobility patterns also contributed to the high accuracy achieved. Real-world implementations may face additional challenges, including multipath propagation, non-Gaussian noise, and more complex jammer movement patterns that may affect the reported performance metrics.

\begin{table}[htbp]
\renewcommand{\arraystretch}{1.0}
\centering
\caption{Summary of Key Performance Metrics}
\label{tab:summary_results}
\begin{tabular}{lc}
\toprule
\textbf{Metric} & \textbf{Value} \\
\midrule
Average SNR Improvement & 9.58 dB \\
Maximum SNR Improvement & 11.08 dB \\
DoA Estimation Accuracy & Up to 99.8\% \\
\midrule
Avg. Processing Time per Collection: & \\
\quad - MUSIC & $\approx$ 41.9 ms \\
\quad - MVDR & $\approx$ 5.2 ms \\
\quad - ML & $\approx$ 16.9 ms \\
\bottomrule
\end{tabular}
\end{table}

\section{Conclusion}
\label{conclusion}

An adaptive beamforming framework combining MUSIC and MVDR was presented and evaluated for mitigating mobile jammers in 5G networks. The integration of a lightweight machine learning model was examined to supplement the core algorithm. The key finding indicates that, while the MUSIC-MVDR baseline provides strong and effective anti-jamming performance, the contribution of a simple linear regression model to overall SNR improvement is marginal. This suggests that significant enhancements in highly dynamic environments require more sophisticated predictive models. Future work will focus on implementing the framework on a hardware testbed using SDR and exploring advanced deep learning architectures to address multi-jammer scenarios and more complex mobility patterns.

\section*{\textcolor{black}{Acknowledgment}}
\noindent
This material is based upon work supported in part by NSF under Awards CNS-2120442 and IIS-2325863, and NTIA under Award No. 51-60-IF007. Any opinions, findings, and conclusions or recommendations expressed in this publication are those of the author(s) and do not necessarily reflect the views of the NSF and NTIA.
\bibliographystyle{IEEEtran}
\bibliography{bib/main}
\end{document}